# Analyzing the Local Basis Set Superposition Error – a case study of CO adsorption on rutile(110)

*Wilke Dononelli[a,b],\* and Thorsten Klüner[a]*

a) Institut für Chemie, Carl von Ossietzky Universität Oldenburg, Germany
b) Department of Physics and Astronomy, Aarhus University, Denmark

\* Correspondence: wilke.dononelli@uni-oldenburg.de

The aim of this letter is to introduce a new method to analyze the local basis set superposition error (LBSSE) using local counterpoise corrections (CP) in order to converge a basis set for a given compound. Using this approach, we are able to define a basis set composition for a CO molecule adsorbed on rutile(110), which can be regarded as complete. In addition to a new LBSSE analysis, adsorption energies of CO on rutile(110) at the MP2 and CCSD level of theory are presented.

**Introduction:**

Nowadays, most of theoretical surface science studies use density functional theory (DFT) to calculate adsorption energies or reaction barriers. In general, DFT works surprisingly well in most cases. Therefore, we also "jumped" on the non-stoppable "DFT train" for most of our studies. Nevertheless, in sense of the famous quote "Finding the right answer for the right reason" it is necessary to benchmark these DFT studies at a higher and especially more systematic level of theory. The ultimate goal or also referred to as "holy grail of quantum chemistry" is full configuration interaction (full CI) in a complete basis set (complete CI). Unfortunately, for systems with more than ~20 electrons this method or even just the full CI is computationally much too demanding to be used on modern high performance computers[1]. Therefore, other high level methods that cover electron correlation are utilized. Two of the most famous ones are i) coupled cluster theory or more precisely CCSD(T) which is referred to as the "gold standard of quantum chemistry" and ii) 2nd order Moller-Plesset perturbation theory (MP2). The later one might not be as accurate, but covers some aspects of electron correlation for a very low price (in sense of computation time).

A problem that most of standard quantum chemical correlation methods share is the computational cost. The problem is much worse than known from text books. In these books, most times just the scaling of correlation methods is mentioned. The formal scaling for some of the previous mentioned methods is $\mathcal{O}(M^l)$ where $M$ is the formal system size and $l = 4$ for Hartee Fock, $l = 5$ for MP2, $l = 6$ for CCSD and $l = 7$ for CCSD(T). Note that this is just a formal textbook scaling. The real scaling will depend on the implementation and will most likely be dependent on the number of occupied and the number of virtual orbitals. Another problem, which is not too often mentioned in textbooks, is that in order to give accurate results for these correlation methods, bigger basis sets need to be considered. Thus, not just $l$ gets bigger, if a higher level is considered, but additionally $M$ has to be increased. Therefore, it might be a clever idea to increase the basis set just at centers, where an increase is necessary and to use for chemical non-interesting parts of the system a smaller basis set. If adsorption of a molecule on a surface is investigated and adsorption energies should be calculated, a common approach is to use a bigger basis set for the adsorbate and the adsorption center and decrease the size of the basis set, when moving away from this center. This way, one hopes to converge the basis set where it is necessary, but have a reduced basis set at regions that are chemically not of interest.

In this letter, we will discuss how well the commonly used approach works for converging the basis set composition for a given system. We will show how one can utilize the counterpoise correction (CP)[2] of Boys and Bernadi to analyze the local BSSE. The CP here is just used in order to achieve convergence of the so-called basis set incompleteness error (BSIE) without using CP in the end. In addition, we will very briefly discuss the method of increments[3-4] and will report MP2 and CCSD adsorption energies for CO on rutile(110).

**Basis set superposition and basis set incompleteness errors**

In general, a complete basis set (CBS) is a basis with an infinite number of one-particle functions. Since we would not be able to compute an infinitely large space, one commonly reduces the size to a finite basis. The error that occurs by introducing the truncation of our space is called basis set incompleteness error (BSIE). The magnitude of the BSIE $E_{BSIE}$ can be defined as the difference of the energy of a system using a complete basis set $E^{CBS}$ or a finite basis set $E^{AB}$.

$$E_{BSIE} = E^{CBS} - E^{AB} \qquad (1.)$$

Nowadays, there are different methods available accounting for the BSIE. Examples are the CBS methods of Petersson and co-workers[5] or Feller[6]. A benefit of calculations of adsorption energies of molecules on surfaces is, that the BSIE is partially accounted for by error cancelation, as briefly mentioned in the introduction. Basis functions that are located further away from the adsorption center do contribute less or at some point even do not contribute to the adsorption energy. Therefore, in a good approximation, the BSIE can be neglected for these centers.

But when considering the interaction of two separated fragments, e.g. an adsorbate A and a substrate B, another basis set related problem comes into play. When both systems interact, the basis functions localized at one fragment contribute to the description of the other fragment and vice versa. Thus, the basis set for the interacting systems is virtually increased.

$$E_{nonCP} = E^{AB}(AB) - E^{A}(A) - E^{B}(B) \quad (2.)$$

There are different methods available that try to compensate for this error, but the widely applied one is the counterpoise correction by Boys and Bernadi[2] which can be written in its geometry independent form as follows:

$$E_{CP} = E^{AB}(AB) - E^{AB}(A) - E^{AB}(B) \quad (3.)$$

$$E_{BSSE} = E_{nonCP} - E_{CP} \quad (4.)$$

Here, the counterpoise corrected energy $E_{CP}$ is calculated in terms of the difference of the entire system in its basis set $E^{AB}(AB)$ and the energies of the fragments in the basis set of the entire system $E^{AB}(A/B)$ in contrast to the non-corrected energy $E_{nonCP}$ where the energy of the fragments in their own basis $E^{A/B}(A/B)$ is considered.

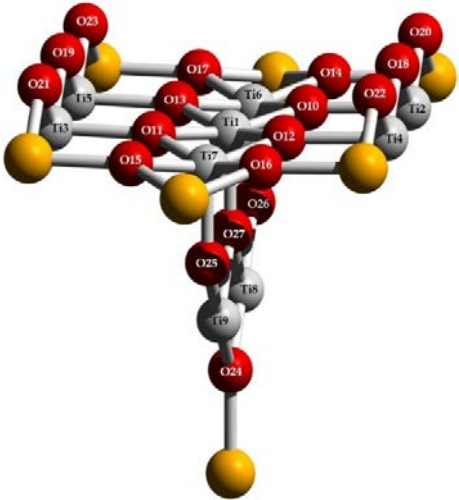

**Figure 1:** $Ti_9O_{18}Mg_7^{14+}$-cluster model of rutile(110). The cluster is embedded in 4000 point charges (not shown).

Although this method is widely used, there is still a debate, whether the CP correction is responsible for introducing other errors[7-8]. In addition, correcting for the BSSE might not completely compensate for the BSIE, and therefore a CP corrected adsorption energy might be as far away from the correct solution as a non-corrected value for too small basis sets. Therefore, the goal of this letter is to present a method where the CP correction is utilized in order to converge to a basis set, where no CP correction is needed.

**Short excursion: The method of local increments**

Before we introduce our scheme how to analyze the local BSSE and utilize this technique to converge our system almost BSIE free, let us first introduce the method of local increments[3-4, 9]. We do so for two reasons. First of all, the method of local increments is a fragment based method utilizing a many-body expansion (MBE). The influence of the BSSE using a MBE has been extensively discussed within the past years[10-12]. The second reason is that this method inspired our analysis method introduced in the next chapter.

The method of increments, first introduced by Stoll[3-4], is a MBE that helps reducing calculation time when high level energies need to be achieved. One way to use the method is to first calculate the energy of the entire system at the HF level of theory $E^{HF}$. Then just the correlation part of the energy $E^{corr}_{MBE}$ is treated using the MBE.

$$E_{tot} = E^{HF} + E^{corr}_{MBE} \quad (5.)$$

$E^{corr}_{MBE}$ is defined in terms of one-body $\varepsilon_i$, two-body $\varepsilon_{ij}$, three-body $\varepsilon_{ijk}$ and higher order contributions according to Equations 6 and 7. A one-body term can be understood as a group of localized orbitals. In substrates, often all orbitals localized at one atom of the surface is chosen as a one-body increment, where it might be useful to define all orbitals localized at the adsorbed molecule as a single one-body increment.

$$\begin{aligned} E^{corr}_{MBE} = & \sum_{i=1}^{N} \Delta\varepsilon_i \\ & + \sum_{i=1}^{N-1}\sum_{j=i+1}^{N} \Delta\varepsilon_{ij} \\ & + \sum_{i=1}^{N-2}\sum_{j=i+1}^{N-1}\sum_{k=j+1}^{N} \Delta\varepsilon_{ijk} \\ & + \cdots \end{aligned} \quad (6.)$$

$$\begin{aligned} \Delta\varepsilon_i &= \varepsilon_i \\ \Delta\varepsilon_{ij} &= \varepsilon_{ij} - \Delta\varepsilon_i - \Delta\varepsilon_j \\ \Delta\varepsilon_{ijk} &= \varepsilon_{ijk} - \Delta\varepsilon_{ij} - \Delta\varepsilon_{ik} - \Delta\varepsilon_{jk} \\ & \quad - \Delta\varepsilon_i - \Delta\varepsilon_j - \Delta\varepsilon_k \end{aligned} \quad (7.)$$

In several publications about adsorption energies of small molecules on oxide surfaces[13-15] it has been mentioned that the method of increments can be truncated either after the second or the third order to achieve adsorption energies that are almost identical to those calculated the conventional way. We tested this for CO adsorbed on rutile(110) which we investigated previously[16-17]. In this underlying work, we found $E_{ads}$ = -1.61 eV ($E_{ads}$(CP) = -0.72 eV) with the method of increments and $E_{ads}$ = -1.65 eV ($E_{ads}$(CP) = -0.73 eV) the conventional way at the MP2 level of theory (see supporting information (SI)). Since the scope of this letter is not to introduce this method, all further computational details can be found in the SI. In addition, we were able to predict the CCSD adsorption energy of CO on rutile(110) with $E_{ads}$ = -1.60 eV ($E_{ads}$(CP) = -0.70 eV) using the method of local increments. All information about the computational setup can be found in the supporting information.

At his point it is worth to mention that it might not be beneficial to use the method of local increments at the MP2 level of theory for system sizes comparable to our example. A conventional calculation will be faster because of the computational scaling.

For equal-sized orbital groups/increments, the method of local increments has a formal scaling of:

$$O\left(\sum_{i=1}^{k} \binom{n}{i} \cdot \left(i \cdot \frac{N}{n}\right)^l\right) \quad (8.)$$

Where $N$ is the abstract system size, $n$ is the number of orbital groups (number of one-body increments), $k$ is the maximal considered order (e.g. if $k=2$ then just one- and 2-body increments will be considered) and $l$ is a correlation method dependent exponential factor. E.g. $l=5$ for MP2, $l=6$ for CCSD or $l=7$ for CCSD(T). Therefore, the method will formally just save computational time, if Equ. 9 is fulfilled.

$$1 - \sum_{i=1}^{k} \binom{n}{i} \cdot \left(\frac{i}{n}\right)^l > 0 \quad (9.)$$

This consideration totally neglects the real scaling of the computational methods, which depend on the implementation. Nevertheless it gives a good starting point for considering weather it is worth using the method of increments or not.

In recent years, the method of increments and additionally the more general MBE were quite frequently used in different studies e.g. adsorption on surfaces[13-15] or interaction of molecules like water monomers[18] and polymers[19-20]. Not all studies show that MBE based methods will by definition work well[10-11], but show up limitations and problem while dealing with solving some of the issues especially the impact of the BSSE[10, 21]. In addition, Manby and co-workers showed how to utilize the method for a two-body energy correction scheme[22-23]. In their method, they do not use the method of increments as a MBE, but utilize two-body increments for energy corrections of lower level computational methods. In a similar fashion, the method of local increments inspired us to create a simple LBSSE analyzation tool as introduced in the next section.

**Local Basis set superposition error analysis**

Similar to conventionally calculated adsorption energies, the ones calculated with the method of local increments will exhibit a BSSE. In order to account for this, the previous mentioned CP correction can be used. We redefine Equation 4 and define the local BSSE energy $\varepsilon_i^{LBSSE}$ (LBSSE) for a given orbital group $i$ as follows:

$$\varepsilon_i^{LBSSE} = \Delta\varepsilon_i^{AB} - \Delta\varepsilon_i^{A/B} \quad (10.)$$

With:

$$\Delta\varepsilon_i^{AB} = \left|\varepsilon_i^{AB}(AB) - \varepsilon_i^{AB}(A/B)\right| \quad (11.)$$

Here $\Delta\varepsilon_i^{AB}$ and $\Delta\varepsilon_i^{A/B}$ are differences between one-body increments for the interacting systems $\varepsilon_i(AB)$ and the non-interacting systems ($\varepsilon_i(A)$ or $\varepsilon_i(B)$) in the basis set of the entire system AB or basis set of the fragments A/B, respectively. One could define a similar error for two- or higher-body increments, but there is no benefit in this. In our study, we define all orbitals localized at a substrate atom or all orbitals localized at the (CO) molecule as a one-body increment. When analyzing the LBSSE, one finds the BSSE for a local atomic environment or local molecular fragment, depending on the definition of the one-body increment or orbital group, respectively. Since quantum chemists are used to define basis functions atom wise in most modern quantum chemistry program packages, this leads to an intrinsic understanding, when also analyzing the LBSSE atom wise. In addition, in the method of local increments (or more general in a MBE) the one-body contribution is the most crucial, since its appearance dominates the expansion (see Equ. 6 and 7). Note, in the case of adsorption of a small molecule like CO on a surface, it might be useful to define all orbitals localized at all atoms of the molecule as one local environment.

**Table 1:** Basis set compositions for different BSSE analysis runs A to F[a)]

| Label | A | B | C | D | E | F |
|---|---|---|---|---|---|---|
| CO | DZ | TZ | TZ | QZ | QZ | 5Z |
| Ti1 | DZ | TZ | TZ | TZ | QZ | 5Z |
| Ti2-5 | DZ | DZ | TZ | TZ | TZ | TZ |
| Ti6-7 | DZ | DZ | DZ | TZ | TZ | TZ |
| Ti8-9 | DZ | DZ | DZ | DZ | TZ | TZ |
| O10-13 | DZ | DZ | DZ | TZ | TZ | QZ |
| O14-17 | DZ | DZ | DZ | DZ | DZ | TZ |
| O18-19 | DZ | DZ | DZ | DZ | DZ | TZ |
| O20-23 | DZ | DZ | DZ | DZ | DZ | DZ |
| O24 | DZ | DZ | DZ | DZ | DZ | DZ |
| O25-26 | DZ | DZ | DZ | DZ | DZ | DZ |
| O27 | DZ | DZ | DZ | DZ | DZ | TZ |
| $E_{ads}$(noCP) | -1.02 | -1.27 | -1.14 | -1.11 | -1.03 | -0.91 |
| $E_{ads}$(CP) | -0.41 | -0.75 | -0.75 | -0.79 | -0.82 | -0.82 |

a) See Figure 1 for atom labels.

**Table 2:** LBSSE energies $\varepsilon_i^{LBSSE}$ for different BSSE analysis runs A to E (Absolute values of energy differences are given in eV)[a)]

| Label | factor | A | B | C | D | E |
|---|---|---|---|---|---|---|
| CO | 1 | 0.100 | 0.037 | 0.039 | 0.015 | 0.016 |
| Ti1 | 1 | 0.078 | 0.017 | 0.019 | 0.034 | 0.019 |
| Ti2-5 | 4 | 0.008 | 0.016 | 0.002 | 0.003 | 0.003 |
| Ti6-7 | 2 | 0.011 | 0.020 | 0.020 | 0.004 | 0.004 |
| Ti8-9 | 2 | 0.008 | 0.012 | 0.011 | 0.018 | 0.002 |
| O10-13 | 4 | 0.003 | 0.013 | 0.011 | 0.006 | 0.005 |
| O14-17 | 4 | 0.000 | 0.000 | 0.000 | 0.002 | 0.003 |
| O18-19 | 2 | 0.002 | 0.004 | 0.002 | 0.004 | 0.005 |
| O20-23 | 4 | 0.001 | 0.002 | 0.001 | 0.002 | 0.002 |
| O24 | 1 | 0.000 | 0.000 | 0.000 | 0.000 | 0.001 |
| O25-26 | 2 | 0.001 | 0.001 | 0.001 | 0.002 | 0.001 |
| O27 | 1 | 0.005 | 0.010 | 0.009 | 0.012 | 0.012 |

In order to test the analyzation method for the local BSSE, we used the previous mentioned system of CO adsorbed on a rutile(110) embedded cluster model. Information about the model and computational setup can be found in the SI. In order to be able to converge to bigger basis set compositions, we used MP2 as correlation method, but similar results could be achieved at other levels of theory, e.g. CCSD(T). Note however, that the MP2 results just reflect the LBSSE at MP2 level. If switching to another level of theory, a new analysis might be necessary.

In Table 1 we listed $\varepsilon_i^{LBSSE}$ values for different basis set compositions. In contrast to our previous studies[16-17] and to the calculations shown in the SI, we used Dunnings correlation consistent basis sets cc-pVXZ (with X = D, T, Q or 5) in order to understand the convergence behavior more rigorously.

For the first basis set composition (all cc-pVDZ), adsorption energies differ by 0.61 eV, depending, whether pure energies or counterpoise corrected values are considered. A standard procedure therefore is to trust in the CP corrected value. But as seen from the biggest basis set composition used in this study, this is not completely converged and will go down from -0.41 eV to -0.82 eV. The non-corrected energies are additionally effected. Here, $E_{ads}$ varies between -1.27 eV to -0.91 eV.

For the basis set composition with least basis functions (A), $E_{ads}$ is in good agreement with the final adsorption energy (F) in the basis set limit, whereas the counterpoise corrected energy shows the biggest error. This result is surprising, but can be explained by a partial compensation of the BSIE by a completely overestimated BSSE. In addition, it is worth to mention that as basis set convergence is not always monotonous but might oscillate, especially for too small basis sets. Nevertheless, after increasing the size of the basis set at the adsorption center (Ti1) and the adsorbate (CO) to triple-$\zeta$ quality, a monotone convergence is achieved from basis set compositions B to F. Information about basis set compositions A to F can be found in Table 1. The corresponding LBSSE values for compositions A to E can be found in Table 2. In the beginning (A) the LBSSE of Ti1 and CO (compare Figure 1) are the largest, at least one order of magnitude larger than the others. Both decrease, when increasing the basis set size, but are still dominant at quadruple $\zeta$ quality (E). As seen from Table 2, sometimes we decided to increase the basis set size, although the LBSSE of a chosen center was smaller compared to another center. But in addition to the pure value it needs to be considered, that these centers a present more often in the material due to the $C_{2v}$ symmetry (presented by the factor given in Table 2). Therefore, these group of symmetrical equivalent centers contribute more to the total BSSE than others.

An obvious result derived from Table 2 is that especially atoms or orbital groups close to the adsorption center exhibit a high LBSSE. This result is neither surprising, nor new, but it is commonly known that a basis set composition has to be chosen in such a way, that the adsorption center is described by most basis functions. Then the number of basis functions may decrease for further distances to the adsorbate. Nevertheless, now we are able to determine the influence of each center individually using the method described in this letter. This way, one can determine, whether a bigger basis set for a specific center needs to be considered, or if it has been chosen large enough. Thereby an unnecessary large amount of basis function can be avoided. This can drastically reduce the computation time, due to the unfortunate scaling of correlation methods with the basis set size.

**Conclusion**

We have defined a way of determining the local basis set superposition error (LBSSE) using localized orbitals at a specific atomic environment or for molecular fragments. Using this analysis, a convergence of the basis set for a specific subsystem in a complex environment can be achieved.

We were able to show that counterpoise (CP) corrected energies for too small basis set compositions might be less correct than non-corrected. On the other hand, as soon as a moderate basis set size is reached, a faster basis set convergence for CP corrected values is achieved than for non-corrected values.

Next to the LBSSE analysis method, we were able to present a CBS MP2 adsorption energy and a CCSD adsorption energy for CO on rutile(110). Here, the CCSD energy has been calculated using the method of local increments.

**Acknowledgement**

Calculations have been performed at the HPC Cluster CARL, located at the University of Oldenburg and funded by the DFG through its Major Research Instrumentation Program (INST 184/157-1 FUGG) and the Ministry of Science and Culture of the Lower Saxony State. We want to thank Carsten Müller, Beate Paulus and Volker Staemmler for introduction to the method of increments.

# Supporting Information:
# Analyzing the Local Basis Set Superposition Error
# – a case study of CO adsorption on rutile(110)

*Wilke Dononelli[a,b],* and Thorsten Klüner[a]*

a) Institut für Chemie, Carl von Ossietzky Universität Oldenburg, Germany
b) Department of Physics and Astronomy, Aarhus University, Denmark

* Correspondence: wilke.dononelli@uni-oldenburg.de

**SI 1: Computational Methods**

All calculations have been performed using Molpro 2010[1] or Gaussian16[2]. All adsorption energies listed in Table 1 of the main text have been calculated at the MP2 level of theory using Gaussian. The adsorption energies listed in SI 2 for our basis set composition used in previous works[3-4] have been calculated in Molpro. Here, first, HF energies have been calculated for the entire system. Then orbitals have been localized using the Foster-Boys method[5-6]. The localized orbitals have been sorted in orbital groups according to the nearest atom center. These orbital groups are so-called one-body increments $\varepsilon_i$ as mentioned in Eq. 6 and 7 in the manuscript. Then, MP2 or CCSD correlation energies have been calculated for all one- and two- body increments, and the total correlation energy has been calculated according to Eq. 6. Note that the LMP2 module of Molpro has been used for MP2 calculations. All thresholds of LMP2 have been chosen in such a way that a LMP2 calculation for the entire system gives exactly the same value as canonical MP2.

**SI 2: MP2 and CCSD adsorption energies for CO adsorbed on rutile(110) calculated with the method of increments**

In two previous works, we presented adsorption energies of CO on rutile(110) at the MP2 level of theory[3-4]. Using the same basis set as presented in this works, we calculated a PES for the linear adsorption/desorption of CO on top of the 5-fold low coordinated Ti atom. Adsorption energies at the HF level and MP2 level have been calculated conventionally. Additionally energies at the MP2 level of theory have been calculated using the method of local increments are shown in Figure S1. All centers of the $Ti_9O_{18}Mg_7^{14+}$-cluster except the $Mg^{2+}$ ions have been considered as one-body increments. We tested to include the $Mg^{2+}$ ions, but no significant change in adsorption energies has been achieved (see Table S1). Since we just considered the linear adsorption of CO in this study, we can make use of the $C_{2v}$ symmetry of the system. Thus, we have to calculate much less incremental energies compared to the entire system. This way, the total number of calculations can be reduced from 406 to 131 for CO on $Ti_9O_{18}$.

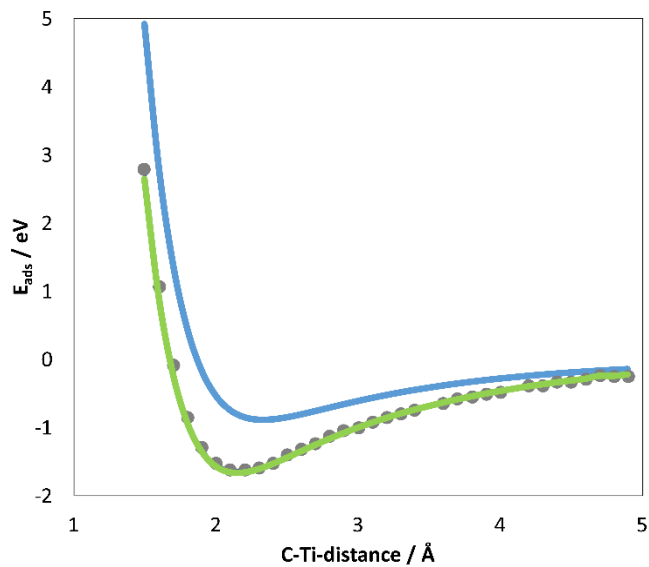

**Figure S1:** Potential energy curve for the linear adsorption of CO on rutile(110). Blue: HF energies; green: conventional MP2; gray dots: MP2 energies calculated using the method of local increments

Figure S1 clearly shows that incremental energies and conventional energies are in very good agreement even when just one- and two- body increments are considered. In Table S1 all counterpoise and not counterpoise corrected CO adsorption energies are listed at both MP2 and CCSD level of theory.

**Table S1:** Adsorption energies of CO adsorbed on an embedded $Ti_9O_{18}Mg_7^{14+}$-cluster

|  | $E_{ads}$ | $E_{ads}(CP)$ |
|---|---|---|
| ***MP2 conv.*** | -1.65 eV | -0.72 eV |
| ***MP2 incr.*** | -1.61 eV | -0.72 eV |
|  | 97.7% | 98.7% |
| ***MP2 incr. (incl. $Mg^{2+}$)*** | -1.61 eV | -0.72 eV |
|  | 97.5% | 98.7% |
| ***CCSD incr.*** | -1.60 eV | -0.70 eV |
| ***CCSD incr. (incl. $Mg^{2+}$)*** | -1.60 eV | -0.70 eV |

Including just one- and two- body increments will result in an error smaller that 2.5% compared to the conventional calculated MP2 energy. An interesting result of this study is that CCSD and MP2 energies are virtually identical. Nevertheless, compared to the basis set analysis presented in the main text, all energies shown in Table S1 seem to be significantly affected by BSIE and BSSE. For this basis set composition, the error for the counterpoise corrected adsorption energy is less than 0.15 eV whereas the non-corrected value overestimates the binding strength by almost 100%.

**Table S2:** LBSSE energies $\varepsilon_i^{LBSSE}$ for BSSE analysis of CO on rutile(110) for MP2 adsorption minimum (Absolute values of energy differences are given in eV)

| Label | $\Delta\Delta\varepsilon_i^{BSSE}$ |
|---|---|
| CO | 0.029 |
| Ti1 | 0.026 |
| Ti2-5 | 0.011 |
| Ti6-7 | 0.018 |
| Ti8-9 | 0.004 |
| *O10-13* | 0.027 |
| *O14-17* | 0.002 |
| *O18-19* | 0.009 |
| *O20-23* | 0.005 |
| *O24* | 0.002 |
| *O25-26* | 0.005 |
| *O27* | 0.059 |

From a LBSSE analysis (see Table S2) as presented in the manuscript it is obvious that besides the basis sets chosen for adsorption center and CO, especially O27 and O10-O13 a poorly described, which results in a large BSSE.